\documentclass{article}

    \usepackage[final,nonatbib]{neurips_2020_ml4ps}

\usepackage[utf8]{inputenc} 
\usepackage[T1]{fontenc}    
\usepackage{hyperref}       
\usepackage{url}            
\usepackage{booktabs}       
\usepackage{amsfonts}       
\usepackage{nicefrac}       
\usepackage{microtype}      
\usepackage{graphicx}

\title{Predicting galaxy spectra from images with hybrid convolutional neural networks}

\author{%
  John~F. Wu\thanks{Alternate email: jwuphysics@gmail.com} \\ Space Telescope Science Institute\\ 3700 San Martin Drive\\ Baltimore, MD 21218 \\ \texttt{jowu@stsci.edu} \\
  \And
  J.~E.~G. Peek\thanks{Secondary affiliation: Department of Physics \& Astronomy, Johns Hopkins University, 3400 N. Charles Street, Baltimore, MD 21218} \\ Space Telescope Science Institute\\ 3700 San Martin Drive\\ Baltimore, MD 21218 \\ \texttt{jegpeek@stsci.edu} \\
}

\begin{document}

\maketitle

\begin{abstract}
Galaxies can be described by features of their optical spectra such as oxygen emission lines, or morphological features such as spiral arms.
Although spectroscopy provides a rich description of the physical processes that govern galaxy evolution, spectroscopic data are observationally expensive to obtain.
For the first time, we are able to robustly predict galaxy spectra directly from broad-band imaging. 
We present a powerful new approach using a hybrid convolutional neural network with deconvolution instead of batch normalization; this hybrid CNN outperforms other models in our tests.
The learned mapping between galaxy imaging and spectra will be transformative for future wide-field surveys, such as with the Vera C. Rubin Observatory and \textit{Nancy Grace Roman Space Telescope}, by multiplying the scientific returns for spectroscopically-limited galaxy samples. 
\end{abstract}

\section{Introduction}

Galaxies are shaped by the physics of their stars, gas, dust, central supermassive black hole, and dark matter.
Our understanding of galaxy formation and evolution hinges on observations of these physical processes, such as the formation of new stars from dense gas clouds, or spectacular mergers between gas-laden massive galaxies.
Spectroscopic observations are necessary for characterizing the stellar populations and interstellar medium of galaxies, e.g., by determining elemental abundances, ionizination state, gas temperature and density, dust properties, and much more [3].

Modern astronomical survey telescopes can quickly scan the sky and capture deep images of galaxies in a few broad-band filters.
While efficient imaging is transforming our view of the Universe, spectroscopic follow-up for individual targets is prohibitively expensive: spectroscopy requires $\sim 10^3$-fold the observation time compared to imaging and cannot be parallelized as easily.
However, optical-wavelength spectroscopy is crucial for investigating the detailed interstellar contents of galaxies. 
Future imaging-only surveys such as the Legacy Survey of Space and Time (LSST) and those with the {\it Nancy Grace Roman Space Telescope} ({\it RST}) will be spectroscopically limited, and their scientific legacies will hinge on how well we can maximize information from the image domain.

Fortunately, deep learning can bridge the gap between prohibitively costly spectroscopy and plentiful photometric imaging.
Spectra can be thought of the labels that are expensive to attain against the cheaper images.
Spectroscopic quantities such as galaxies' elemental abundances or gas/stellar mass ratios can already be estimated directly from optical imaging [9,10].

\begin{figure}
  \centering
  \includegraphics[width=0.99\textwidth]{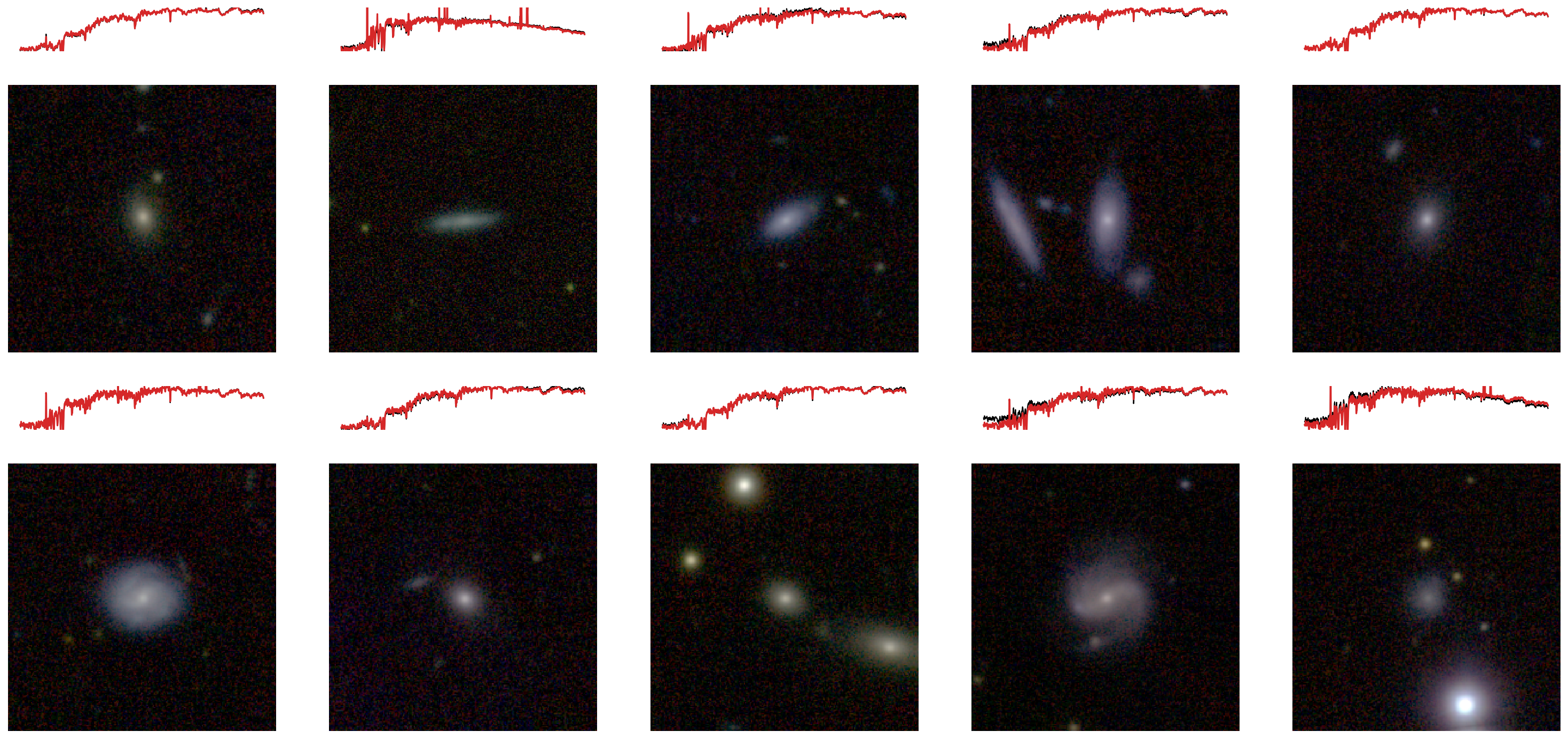}
  \caption{
    Randomly selected examples of PanSTARRS $grizy$ image inputs and reconstructed galaxy spectrum outputs. 
    VAE-decoded targets and predictions are shown in black and red, respectively.
    \label{fig:result}
    }
\end{figure}

In this work, we predict galaxy spectra solely from color images.
Specifically, we train a deep convolutional neural network (CNN) to map five-channel Pan-STARRS image cutouts of galaxies onto latent space representations of their SDSS spectra.
We successfully reconstruct observed galaxy spectra directly from images.
Examples of the galaxy image inputs and spectra outputs from our model are shown in Figure~\ref{fig:result}.
Our method is broadly applicable in astronomy and can be used for generating predictions, pre-selecting interesting candidates for spectroscopy, and interpreting morphological connections to spectral features.
The code is available online at \url{https://github.com/jwuphysics/predicting-spectra-from-images}.

\section{Method}

\begin{figure}
  \centering
  \includegraphics[width=0.95\textwidth]{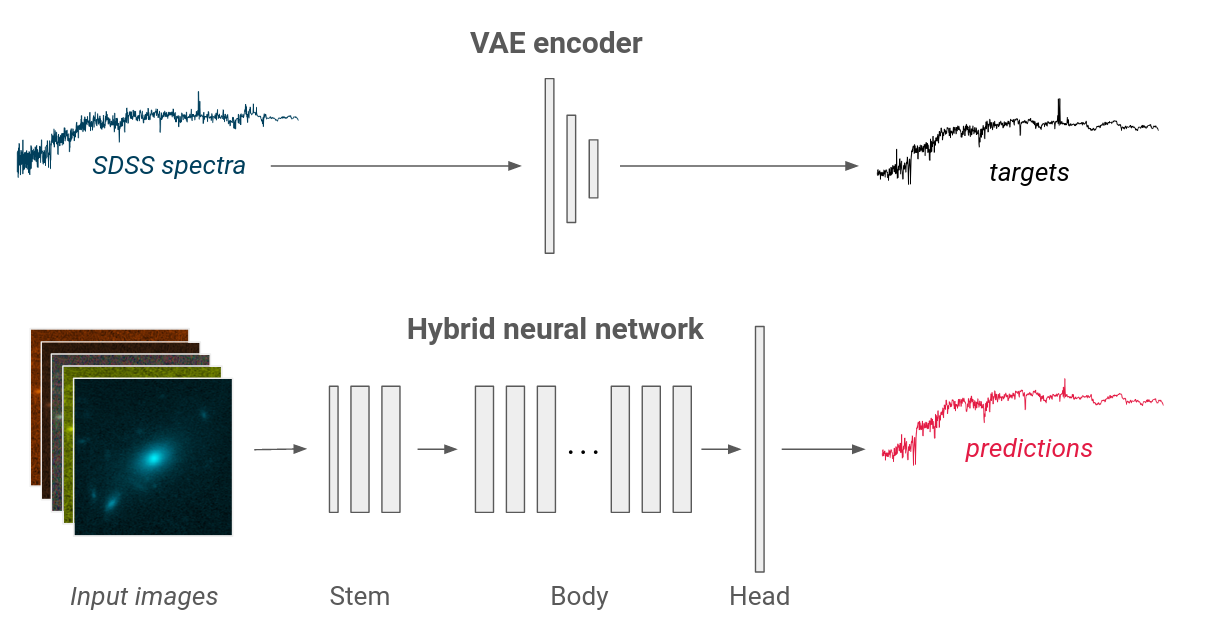}
  \caption{A schematic of our methodology.
  A pretrained VAE maps SDSS spectra to six latent variables, which we use as training targets ({\it upper}). 
  We optimize a CNN to estimate the latent variables from $grizy$ galaxy images ({\it lower}).
  Our best model comprises a deconvolution stem, resnet-like CNN body, and fully-connected layer head.
  While the loss function compares targets and predictions in the six-dimensional latent space, we show examples of the decoded spectra for visual comparison.
  \label{fig:method}
  }
\end{figure}

In \S2.1, we summarize a variational autoencoder (VAE) method for reconstructing a galaxy's optical spectrum from six latent variables.
In \S2.2, we describe Pan-STARRS 1 $grizy$ (five-band) galaxy image cutouts that serve as CNN inputs.
In \S2.3, we train a CNN to predict these latent variables, and compare traditional CNN architectures against recently proposed models, which use \textit{deconvolution} layers for feature normalization instead of batch normalization (batchnorm).
A schematic representation of the method is shown in Figure~\ref{fig:method}.

\subsection{Latent space targets}

Following Portillo et al. [8], we use a catalog of 64,000 objects that have been targeted by Sloan Digital Sky Survey (SDSS) fiber spectroscopy, the majority of which are representative galaxies from the SDSS main galaxy sample.
Their optical spectra are shifted to a common velocity frame, normalized, and re-sampled onto a grid of 1000 elements.
After quasars and other objects are removed, the remaining catalog is split using 80\% for training and 20\% for validation.

We employ a trained VAE to represent these optical-wavelength spectra using six latent variables, which correspond to star formation properties, emission-line ratios, post-starburst state, nuclear activity, and other global galaxy properties [8].
VAEs are composed of an encoder portion, which maps galaxy spectra to the latent space, and a decoder, which performs the inverse mapping from latent variables to spectra.
Other works have used principal components analysis (PCA) or non-negative matrix factorization (NMF) for dimensionality reduction, but VAEs are better at representing complex spectral features and non-linear connections between features.
Moreover, VAEs map these features onto a continuous latent space, which ensures that reconstructed galaxy spectra can smoothly interpolate between examples.

\subsection{Galaxy image cutouts}

We have obtained images of galaxies in five broad-band filters ($grizy$) from Data Release 2 of the Pan-STARRS 1 survey (\url{https://panstarrs.stsci.edu/}).
The $224 \times 224 $ image cutouts are delivered in \texttt{FITS} format with an angular scale of $0.25^{\prime\prime} ~{\rm pixel}^{-1}$.
We augment the data using D$_4$ dihedral group transformations.
Most images have other astronomical objects in them, which may convey details about the galaxy environment (or inject irrelevant information about background or foreground sources).
Although a small number of cutouts have imaging artifacts, we do not attempt to remove them in this work.

\subsection{Network deconvolution}

CNNs are supervised machine learning algorithms that can encapsulate detailed representations of images.
In recent years, CNNs have become widely used for astronomical classification and multivariate regression tasks, in addition to other computer vision problems [1,7].
A recently proposed architecture modification introduces \textit{deconvolution} layers [11], which remove pixel-wise and channel-wise correlations from input features.
Network deconvolution allows for efficient optimization and sparse representation via convolutional kernels.

\section{Results}

\subsection{Experiments}

We compare CNN models with different feature normalization methods.
The base architecture is an 18-layer residual neural network (resnet) modified with several enhancements, including self-attention layers, Mish activation functions, and a modified stem comprising only $3 \times 3$ convolutions.
It has previously been shown that these modifications improve the commonly used 18-layer resnet for astronomical regression and classification tasks [10].
We test models containing the usual batchnorm layers (xresnet18), deconvolution in lieu of batchnorm throughout the model (xresnet18-deconv), and deconvolution in the CNN stem and batchnorm in the body (xresnet18-hybrid). 
We also test a baseline model composed of three convolution layers + batchnorm (with 32, 32, and 64 filters), with ReLU activations, followed by a fully connected layer (simple-cnn), and a baseline model with deconvolution layers (simple-deconv).

We optimize the CNN by minimizing the mean squared error (MSE) between latent space predictions and targets. 
Because the VAE is able to capture physically important correlations between spectral features, MSE in the latent space is a desirable similarity metric compared to, e.g., Euclidean distance between decoded spectra.
Learning rates are scheduled according to the Fast.ai one-cycle defaults [2], with a maximum learning rate of 0.03 (except xresnet18-deconv, for which we find an optimal value of 0.003).
We set batch size to 128 and use a combined RAdam [4] and LookAhead [12] optimizer with weight decay of 0.01.
All CNNs are initialized with Kaiming initialization.
These hyperparameter choices have been shown to produce excellent optimization results for xresnet models [10].
In Table~\ref{tab:results}, we report the validation loss after 10 (all models) and 50 training epochs (xresnet18-like models). 
Because latent variables are distributed according to a multivariate Gaussian with unity variance, we can achieve a baseline MSE loss of 1 by always predicting the mean latent vector.

\subsection{Comparing network deconvolution against batch normalization}

The xresnet18-hybrid model performs best in our experiments, implying that deconvolution is most useful in the first few layers of a CNN.
We also find that the xresnet18 with batchnorm outperforms the variant with deconvolution throughout (xresnet18-deconv).
As expected, deeper networks strongly improve the loss relative to simple baseline models, even with a short training duration.

These experiments confirm that network deconvolution promotes efficient optimization.
After 50 epochs, the xresnet18-deconv suffers from heavy overfitting, whereas the xresnet18 and xresnet18-hybrid maintain robust performance.
Overfitting likely occurs because galaxies have relatively simple structures that can be represented in a small number of convolution + deconvolution layer pairs.
We note that our hybrid model may not perform well on more complex objects in the ImageNet or Cityscapes data sets, which may require deep CNNs with additional deconvolution layers [11].

\begin{table}
  \caption{Experiments: CNNs with batchnorm and deconvolution}
  \label{tab:results}
  \centering
  \begin{tabular}{l r r r}
    \toprule
    Model                   & Parameters  & \multicolumn{2}{c}{MSE Loss} \\
                                            \cmidrule(r){3-4}
                            &             & 10 epochs       & 50 epochs   \\
    \midrule
    {\bf xresnet18-hybrid}  & {\bf 6.73M} & {\bf 0.506} & {\bf 0.474} \\
    xresnet18               & 6.73M & 0.528           & 0.487         \\
    xresnet18-deconv        & 6.73M & 0.541           & 0.577         \\
    \cmidrule(r){1-4}
    simple-cnn              & 29.7k & 0.783           &  -    \\
    simple-deconv           & 29.6k & 0.678           &  -    \\
    \bottomrule
  \end{tabular}
\end{table}

\section{Discussion}

For the first time, we are able to predict a galaxy's spectrum directly from imaging.
We posit that the hybrid approach of using deconvolution plus convolution layers in the stem and convolution plus batchnorm layers throughout the rest of the neural network encourages the model to represent images sparsely with independent low-level features, while permitting the subsequent layers to learn redundant features necessary for expressing symmetries in the data, such as rotation, translation, and scale equivariance.
Our hybrid model can be useful in other disciplines that leverage computer vision, and may be particularly valuable for highly correlated input data (e.g., jet identification in quark-gluon plasma, or remote sensing with interferometric aperture synthesis).

The robust mapping between galaxy images and spectra implies that the appearance of a galaxy is a {\it strong prior} on its spectrum.
Since the VAE ensures that images are projected onto a smoothly varying low-dimensional latent space, we can explicitly estimate these priors by using Bayesian neural networks or Monte Carlo dropout in the head of our neural networks.
These image-domain priors will be critical for selecting galaxies with particular spectral features in targeted follow-up studies.
It is important to note that our understanding of astronomy is often challenged and advanced by discovery of rare phenomena.
For this reason, our deep learning approach will also be useful for identifying new failure modes, e.g., galaxies which have aberrant spectra given their morphology, and for interpreting the physical connections between galaxy imaging and spectra [5,10].

\section*{Broader Impact}

Let us consider the negatives first.
Computer vision, empowered by deep learning, has led to harmful outcomes in warfare and mass surveillance, and has propagated or reinforced deleterious biases in medicine and image recognition.
Marginalized communities are disproportionately impacted, and our work may contribute to these disparities. 
However, advancements in analyzing astronomical data may not translate to similar gains (and thus, potential for abuse) in other applications, particularly because the information in galaxy images is fundamentally different from the contents of terrestrial images.
In terms of positive broader impacts, our hybrid CNN architecture results in improved optimization, which is helpful for reducing the deep learning carbon footprint.
By predicting priors on spectra from existing galaxy images, our method can promote more efficient use of telescope time.
Finally, we believe that our work on open data sets encourages development of community tools and facilitates accessible science.
There is strong evidence that robust scientific archives increase accessibility, which benefits junior, under-resourced, and diverse researchers from a broader set of institutions [6].

\begin{ack}

John Wu acknowledges support from the National Science Foundation under grants NSF AST-1517908 and NSF AST-1616177.
This work was supported by Google Cloud Platform research credits. 
\end{ack}

\section*{References}

\medskip

\small

\bibliography{bibliography.bib}

[1] Dieleman, S., Willett, K.~W., \& Dambre, J.\ (2015)  Rotation-invariant convolutional neural networks for galaxy morphology prediction, {\it Monthly Notices of the Royal Astronomical Society}, 450, 1441--1459.

[2] Howard, J., Gugger, S. (2020) Fastai: A Layered API for Deep Learning, \textit{Information}, 11, 108--134. 

[3] Kewley L.J., Nicholls D.C., Sutherland R.S. (2019)  Understanding Galaxy Evolution Through Emission Lines, {\it Annual Review of Astronomy and Astrophysics}, 57, 511--570.

[4] Liu, L., Jiang, H., He, P., Chen, W., Liu, X., Gao, J., Han, J. (2020) On the Variance of the Adaptive Learning Rate and Beyond, {\it 8th International Conference on Learning Representations}, 13 pp.

[5] Peek, J.E.G. \& Burkhart, B.\ (2019),  Do Androids Dream of Magnetic Fields? Using Neural Networks to Interpret the Turbulent Interstellar Medium, {\it The Astrophysical Journal Letters}, 882, 12--20.

[6] Peek, J., Desai, V, White, R.L., D'Abrusco, R., Mazzarella, J.M., Grant, C., Novacescu, J., Scire, E., Winkelman, S. (2019) {\it Astro2020: Decadal Survey on Astronomy and Astrophysics}, 51, 105, 10 pp.

[7] Peek, J.E.G.,  Jones, C.K., Hargis, J. (2020) Convolutional Neural Networks in Astronomy, and Applications for Diffuse Structure Discovery, {\it Astronomical Data Analysis Software and Systems XXVII}, 522, 381--385.

[8] Portillo, S.K.N., Parejko, J.K., Vergara, J.R., Connolly, A.J. (2020) Dimensionality Reduction of SDSS Spectra with Variational Autoencoders, {\it The Astronomical Journal}, 160, 45--62.

[9] Wu, J.F. \& Boada, S. (2019)  Using convolutional neural networks to predict galaxy metallicity from three-colour images, {\it Monthly Notices of the Royal Astronomical Society}, 484, 4683--4694.

[10] Wu, J.F. (2020) Connecting Optical Morphology, Environment, and H\,{\sc i} Mass Fraction for Low-Redshift Galaxies Using Deep Learning, {\it The Astrophysical Journal}, 900, 142--160.

[11] Ye, C., Evanusa M., He, H., Mitrokhin, A., Goldstein, T., Yorke, J.A., Ferm{\"u}ller, C., Aloimonos, Y. (2020) Network Deconvolution, {\it 8th International Conference on Learning Representations}, 20 pp.

[12] Zhang, M.R., Lucas, J. and Hinton, G. and Ba, J. (2019) Lookahead Optimizer: k steps forward, 1 step back, {\it Advances in Neural Information Processing Systems 32}, 9597--9608.

\end{document}